\documentclass[reprint,prl,twocolumn]{revtex4-2}

\usepackage{graphicx}
\usepackage{amsfonts}
\usepackage{amsmath}
\usepackage[usenames,dvipsnames]{color}
\usepackage{bm}
\usepackage{amssymb}
\usepackage{braket}

\usepackage[usenames,dvipsnames]{color}

\usepackage{multirow}
\usepackage{epstopdf}
\usepackage[normalem]{ulem}

\usepackage{hyperref}
\hypersetup{
colorlinks=true,
linkcolor=blue,
citecolor=blue,
filecolor=green,
urlcolor=blue,
}

\begin{document}

\title{Robust nuclear spin entanglement via dipolar interactions in polar molecules}
\date{\today}

\author{Timur V. Tscherbul$^{1}$, Jun Ye$^{2}$, and Ana Maria Rey$^{2}$}

\affiliation{$^{1}$Department of Physics, University of Nevada, Reno, Nevada, 89557, USA}
\affiliation{$^{2}$JILA, National Institute of Standards and Technology, and Department of Physics, University of Colorado,  Boulder, Colorado, 80309, USA}

\date{\today}

\begin{abstract}
We propose a general protocol for on-demand generation of robust entangled states of nuclear and/or electron spins of ultracold $^1\Sigma$ and $^2\Sigma$ polar molecules  using electric dipolar interactions. By encoding a spin-1/2 degree of freedom in a combined set of spin and  rotational molecular levels, we theoretically demonstrate the emergence of effective spin-spin interactions of the Ising and XXZ forms, enabled by efficient magnetic control over electric dipolar interactions. We show how to use these interactions to create long-lived cluster and squeezed spin states.
\end{abstract}

\maketitle

Ultracold polar molecules hold great promise for {quantum simulation, metrology and information processing} because they feature strong {electric dipolar (ED)} interactions that are both long range, anisotropic, and, more importantly, tunable  \cite{DeMille:02,Andre:06,Yelin:06,Wei:11,Karra:16,Ni:18,Yu:19,Wall:09,Carr:09,Gorshkov:11a,Gorshkov:11b,Yan:13,Bohn:17,Yao:18,Albert:20,Kaufman:21}.  A necessary condition towards their use for these  goals is the capability  to  take advantage of their intrinsic  ED interactions  to create highly entangled and long-lived molecular states that are robust to environmental decoherence, {such as spin-squeezed states for enhanced sensing \cite{Ma:11,Pezze:18,Bilitewski:21}}, or {cluster states for measurement-based} quantum computation \cite{Briegel:01,Raussendorf:03,Briegel:09,Lanyon:13,Mamaev:19,Kuznetsova:12}. 

Up to date, rotational states of simple bialkali molecules such as KRb have been proposed as the primary workhorse and a natural degree of freedom to encode a qubit \cite{DeMille:02,Andre:06,Yelin:06,Wei:11,Karra:16,Ni:18,Yu:19,Wall:09,Carr:09,Gorshkov:11a,Gorshkov:11b,Yan:13}. This is because the long-lived rotational states can be directly coupled by long-range ED interactions and manipulated by microwave (mw) fields \cite{Ospelkaus:10a,Blackmore:20b}. 
Nevertheless, rotational states feature important limitations which have hindered their use for entanglement generation: (1) ultracold molecules prepared in different rotational states typically experience different trapping potentials and therefore are subject to undesirable decoherence, leading to short coherence times \cite{Kotochigova:10b,Caldwell:20,Burchesky:21}; (2) fine tuning of the many-body Hamiltonian parameters
requires the use of strong and well-controlled dc electric fields $E$ \cite{DeMille:02,Gorshkov:11b}. As these fields take time to switch and change,  on-demand generation of {entanglement using} long-range ED interactions between rotational states remains a significant  experimental challenge.

To overcome these important limitations, here we propose to leverage a larger set  of  internal levels accessible in ultracold polar molecules, which include nuclear and/or electron spin levels in addition to their rotational structure. Taken together, these levels can be used as  a robust resource for on-demand entanglement generation. By encoding an effective spin-1/2 into a combined set of  nuclear spin and  rotational molecular levels, we take advantage of the long coherence times enjoyed by nuclear spin levels and  the strong ED interactions  experienced by  rotational levels at different stages of the entanglement generation process. 
We note that other types of qubit encodings have been suggested in Refs.~\cite{Wei:11,Karra:16}.
 Our proposal, in addition to the electric tunability of  dipole-induced  spin-exchange and Ising couplings,  offers magnetic tunability of the ED interactions, and the ability to turn off exchange interactions on demand even at finite $E$-fields. The magnetic tunability, an essential advantage of our approach,  arises from inducing avoided crossings of electron spin-rotational levels of the opposite parity  in $^2\Sigma$ molecules such as
YO \cite{Collopy:15,Collopy:18,Ding:20,Wu:21}, CaF \cite{Truppe:17,Anderegg:18}, and SrF \cite{Barry:14,McCarron:18}, which have already been cooled down and trapped in {optical potentials}. 
We also show how to engineer magnetically tunable level crossings in $^1\Sigma$ molecules via mw dressing.
In this way, we can engineer spin models of Ising and XXZ types  with an enlarged set of control parameters.


\begin{figure}[t]
\includegraphics[scale=0.79,trim = 0 0 0 0]{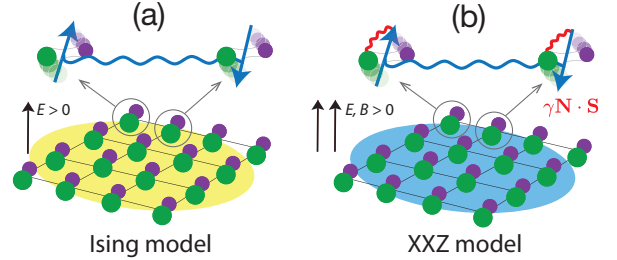}
\caption{(a) Experimental setup for nuclear spin entanglement generation consisting of  $^1\Sigma$ molecules in superpositions of spin-rotational states trapped {in an optical lattice on a 2D plane}. 
Molecular spins (arrows) interact via the long-range Ising interaction (wavy line), creating cluster-state entanglement represented by the yellow shaded area.
(b) Same as (a) but in the presence of a $B$-field and the spin-rotation interaction (red wavy lines) near an avoided crossing in $^2\Sigma$ molecules. The blue shaded area represents entanglement of the XXZ type leading to the generation of spin-squeezed states.}
\label{fig:cartoon}
\end{figure}

{\it Entangling nuclear spins of $^1\Sigma$ molecules: Ising interactions and cluster states}. Consider an ensemble of ultracold $^1\Sigma$  molecules   {confined to a single plane of a three-dimensional (3D)} optical lattice at unit filling as shown in Fig.~\ref{fig:cartoon}(a). To entangle the nuclear spins of  the molecules via ED interactions, we propose to encode an effective spin-1/2 into the states
 $\ket{\uparrow} = \ket{\tilde{0}0,M}$ and $\ket{\downarrow} =  \ket{\tilde{1}0,M'}$ as shown in Fig.~\ref{fig:Ising}(a). {We refer to these states as nuclear spin-rotational states.}  Here,  $|\tilde{N}M_N, M\rangle$ denote molecular eigenstates in the limit of large magnetic field $B$, where  the nuclear spin projections on the field axis $M=\{M_{I_1},M_{I_2}\}$ are good quantum numbers \cite{Aldegunde:08,Ospelkaus:10a}.
  At zero $E$-field the rotational quantum number $N$ is a good quantum number with {the rotational} energy spacing set by the rotational constant $B_e$. At finite $E$-field, the rotational states admix but we can  still label the  states that adiabatically connect to {zero-field $N$-states} as $\tilde{N}$ \cite{Wall:15c}.
 We further assume that the external $E$ and $B$-fields are parallel.
 

 Importantly, we require that our effective spin-1/2 states have {\it different} nuclear spin projections ($M\ne M'$), which ensures that the off-diagonal matrix elements of the molecular electric dipole moment (EDM) $\hat{d}$ are strongly suppressed ($d_{\uparrow\downarrow}\to 0$).
This is because in the large $B$-field limit,  the electric quadrupole interaction, 
becomes small compared to the Zeeman interaction \cite{SM}, and $M$ and $M_N$ become good quantum numbers.
 This property makes our encoding distinct from the one proposed in all of the previous theoretical work, which used rotational states with $M=M'$ \cite{DeMille:02,Yelin:06,Andre:06,Wall:09,Wei:11,Ni:18,Yu:19,Gorshkov:11b}. The only exception is Ref.~\cite{Herrera:14}, which proposed the use of  {a combination of} electron spin and rotational  states near avoided level crossings to engineer interqubit interactions using infrared laser pulses. In our work, these interactions arise naturally without the need to use extra pulses or avoided crossings (except  in the specific case of the XXZ model with $^1\Sigma$ molecules as discussed below).
 

In the absence of an external $E$-field, all matrix elements of the EDM vanish identically in our effective spin-1/2 basis [see Fig.~\ref{fig:Ising}(a)], as the basis states have a definite parity. As a result, the long-range ED interaction between the molecules also vanishes.
When an $E$-field is applied, the diagonal matrix elements  $d_{\uparrow} = \bra{\uparrow}  \hat{d}  \ket{\uparrow}$ and  $d_{\downarrow} = \bra{\downarrow}  \hat{d}  \ket{\downarrow}$ acquire finite values with $d_{\uparrow}\ne d_{\downarrow}$, as shown in  Fig.~\ref{fig:Ising}(b), whereas  $d_{\uparrow\downarrow}\to 0$ (see above).
This leads to the emergence of ED interactions between molecules in the different nuclear spin-rotational states described  by the long-range Ising model \cite{SM}
\begin{equation}\label{Hmb_Ising}
\hat{H}_\text{dip} = \sum_{i>j} J_{ij}^z \hat{S}_i^z\hat{S}_{j}^z,
\end{equation}
where the effective spin-1/2 operators $\hat{S}_i^z$ act in the two-dimensional Hilbert space of the $i$-th molecule $\{{\ket{\downarrow},\ket{\uparrow}} \}$, $J^{z}_{ij}= \frac{1{-}3\cos^2\theta_{ij}}{|\mathbf{R}_{ij}|^3}  (d_{\uparrow} - d_{\downarrow})^2$ is the Ising coupling constant, $\mathbf{R}_{ij}$ is  the distance vector between the molecular centers of mass,  and $\theta_{ij}$ is the angle between $\mathbf{R}_{ij}$ and the direction of the $E$-field.  The $E$-field dependence of $J^{z}_{ij}$  shown in Fig.~\ref{fig:Ising}(b)  reaches a maximum at $\beta_E=Ed/B_e \simeq 3$.

\begin{figure}[t]
\includegraphics[scale=0.5,trim = 0 0 0  0]{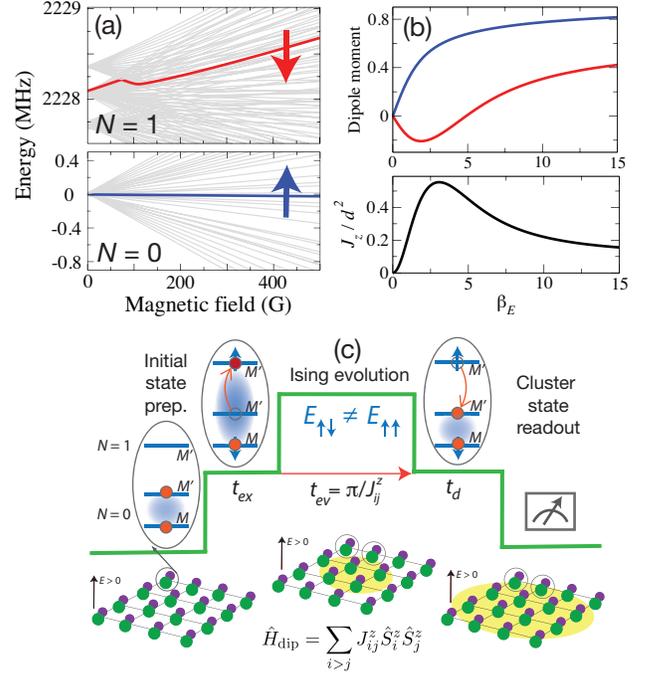}
\caption{(a) Zeeman energy levels of a typical bialkali molecule (here, $^{40}$K$^{87}$Rb).  Our effective spin-1/2 states $\ket{\uparrow} = \ket{\tilde{0}0,M}$ and $\ket{\downarrow} =  \ket{\tilde{1}0,M'}$ are marked by arrows. (b) Expectation values of the EDM $d_\uparrow$ (top curve) and $d_\downarrow$ (bottom curve) and the Ising coupling constant $J_z$ (bottom panel) as a function of the reduced $E$-field strength $\beta_E=dE/B_e$ \cite{Wall:15c}.
(c) Protocol for creating long-lived nuclear spin cluster states of polar molecules via ED interactions.}
\label{fig:Ising}
\end{figure}

 Importantly, because our {effective spin-1/2} states correspond to the $\tilde{N}=0$ and $\tilde{N}=1$ Stark levels,  the Ising interaction between the molecules in different nuclear spin-rotational states {($M\ne M'$) is as strong as that between the molecules in ordinary rotational states  {($M=M'$)}  \cite{Gorshkov:11a}.}


The starting point of our general entanglement-generating protocol is an  ensemble of $\tilde{N}=0$ molecules {trapped in a single plane of a 3D optical lattice} as shown in Fig.~\ref{fig:Ising}(c).  The {molecules are initialized  in a coherent superposition of two different $\tilde{N}=0$ nuclear spin states $|\tilde{0}0,+\rangle = \frac{1}{\sqrt{2}}[\ket{\tilde{0}0,M} + \ket{\tilde{0}0, M’}]$}.  
 No long-range interactions are initially present between the molecules, since $d_\uparrow=d_\downarrow$ for all $\tilde{N}=0$ nuclear spin states even in the presence of a dc $E$-field, and thus $J^z_{ij}=0$.
   As an example, we consider ultracold KRb$(^1\Sigma^+)$ molecules prepared in a coherent superposition of two nuclear spin states {$\ket{\tilde{0}0,+}_\text{KRb} =  \frac{1}{\sqrt{2}}[\ket{\tilde{0}0,-3,-\frac{1}{2}} + \ket{\tilde{0}0,-4,\frac{1}{2}}$]}, which can be realized experimentally via  two-photon microwave excitation  \cite{Ospelkaus:10a,Park:17}.   
 {Further analysis of the experimental feasibility of our protocol \cite{SM} shows that most of the requisite experimental tools have already been implemented in the laboratory, and thus its experimental realization appears feasible with current or near-future  technology.}
   
\begin{figure}[t]
\includegraphics[scale=0.46,trim = 0 0 0  0]{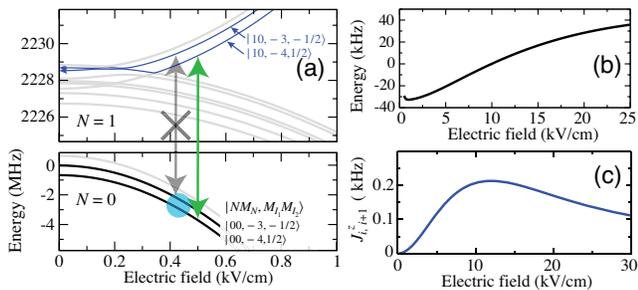}
\caption{(a) Energy levels of $^{40}$K$^{87}$Rb used in our entanglement generation protocol plotted vs $E$-field at $B=400$~G and $M_F=M_N+M_{I_\text{K}}+M_{I_\text{Rb}}=-7/2$. The initial coherent superposition of the $\tilde{N}=0$ nuclear spin states is marked with a blue circle. 
The mw transition used to excite the initial coherent superposition $\ket{+}$ is marked with a green arrow.  The  competing mw transition is marked by the grey arrow.  The differential Stark shift (b) and the NN Ising constant (c) plotted vs $E$-field for $^{40}$K$^{87}$Rb molecules at the lattice spacing of 500~nm.}
\label{fig:IsingKRb}
\end{figure}
 
In {\it Step 1}, we initialize our spin-rotational qubits by preparing a coherent superposition  $\ket{+}=\frac{1}{\sqrt{2}}[\ket{\uparrow} + \ket{\downarrow}]$ [see~Fig.~\ref{fig:Ising}(c)]. This can be achieved by starting with the coherent superposition of $\tilde{N}=0$ nuclear spin states {$|\tilde{0}0,+\rangle$} defined above, and applying a resonant pulse of mw radiation on the $\ket{\tilde{0}0, M’} \to \ket{\tilde{1}0, M’}$ rotational transition.

Figure~\ref{fig:IsingKRb}(a) illustrates the idea for KRb. By applying a mw $\pi$  pulse to the superposition {$\ket{\tilde{0}0,+}_\text{KRb}$} resonant  on the $\ket{\tilde{0}0,-4,\frac{1}{2}}  \to \ket{\tilde{1}0,-4,\frac{1}{2}}$ transition (which has a large transition dipole moment), the population in the $\ket{\tilde{0}0,-4,\frac{1}{2}}$ state is coherently transferred to the rotationally excited state, leading to the desired  superposition  $\ket{+}_\text{KRb}=\frac{1}{\sqrt{2}}[ \ket{\tilde{0}0,-3,-\frac{1}{2}} +  \ket{\tilde{1}0,-4,\frac{1}{2}}]$. To show that it is possible to selectively excite a single eigenstate component of the initial superposition, (i.e, $\ket{\tilde{0}0,-4,\frac{1}{2}}$ but not $\ket{\tilde{0}0,-3,-\frac{1}{2}}$), we plot in  Fig.~\ref{fig:IsingKRb}(b) the energy difference between the main transition ($\ket{\tilde{0}0,-4,\frac{1}{2}}\leftrightarrow \ket{\tilde{1}0,-4,\frac{1}{2}}$) and the {competing} transition.
The differential Stark shift is seen to be negative at small $E$-fields, approaching zero at $E\simeq 10$~kV/cm. 
Thus, selective mw excitation of the desired transition $\ket{\tilde{0}0,-4,\frac{1}{2}}\to \ket{\tilde{1}0,-4,\frac{1}{2}}$ is possible by tuning the $E$-field below  5~kV/cm or above  15~kV/cm, where the {competing} transition is energetically detuned. 

In {\it Step 2}, we let our initial $n$-molecule superposition  {$\ket{+}^{\otimes n}$} evolve under the {Ising} interaction for the cluster time {$t_c=\pi/|J^{z}_{i,i+1}|$. Because of the long-range nature of the interaction, the resulting proxy cluster state will be different from the proper cluster state formed  under nearest-neighbor (NN) interactions \cite{Briegel:01,Raussendorf:03}.}  Nevertheless, theoretical simulations in 1D show that cluster-state fidelities above 75\% can be achieved for ${\leq}~12$ molecules even in the presence of moderately strong $(|J^{z}_{i,i+2}/J^{z}_{i,i+1}|=0.1)$ next-NN interactions \cite{Zhang:11}. These interactions can also be  efficiently suppressed using dynamical decoupling techniques \cite{Zhang:11}. 

At $t=t_c$ a maximally entangled  cluster state of nuclear spin-rotational qubits is created \cite{Briegel:01,Mamaev:19}.
 The Ising coupling constant  $J^z_{i,i+1}$ is  plotted for the NN interactions of KRb molecules in Fig.~\ref{fig:IsingKRb}(c).  At $E=20$~kV/cm and the NN spacing of 500~nm, $J^z_{i,i+1}/2\pi=170$~Hz and it takes $t_c \simeq 2.95$~ms to evolve the lattice-confined ensemble of KRb molecules  into the highly entangled cluster state. 
Preserving the coherence  during Ising evolution is experimentally feasible because the coherence times $T_2$ of the nuclear spin-rotational superpositions are likely to be similar to those of purely rotational superpositions \cite{Park:17,Gregory:21}, with $T_2=470$~ms   demonstrated for CaF \cite{Burchesky:21}. 
{A detailed analysis of the effects of decoherence on cluster states in presented in the Supplemental Material \cite{SM}}.



In {\it Step 3}, we coherently transfer population back to the  ground {rotational state via the $|\tilde{1}0,M’\rangle \to |\tilde{0}0,M'\rangle$} transition. 
 After the deexcitation step, the molecules find themselves once again in a coherent superposition of $\tilde{N}=0$ nuclear spin sublevels $|\tilde{0}0,+\rangle = \frac{1}{\sqrt{2}}[\ket{\tilde{0}0,M} + \ket{\tilde{0}0, M’}]$.  The long-range Ising interaction is thereby completely turned off, preserving the entanglement created in {\it Step~2}. 
We expect the resultant nuclear spin cluster state to be both long-lived and robust due to the much longer  coherence times of $\tilde{N} = 0$ nuclear spin qubits compared to their rotational counterparts \cite{Park:17,Gregory:21}.
These advantages of our protocol enable long-term storage of quantum information in the cluster state encoded in molecular nuclear spins, enabling efficient implementation of the measurement protocols of one-way quantum computing \cite{Raussendorf:03,Briegel:09}.

    We next consider the question of how to engineer a more general XXZ-type interaction between the electron spins of $^2\Sigma$ molecules or nuclear spins of  $^1\Sigma$ molecules.
    While this interaction arises naturally between the different rotational states with $M=M'$ \cite{Gorshkov:11b,Wall:15c}, it requires  finite  off-diagonal  EDM matrix elements, which are absent in the basis of nuclear spin-rotation states we have considered so far.  
    Thus, in order to obtain a nonzero spin exchange  coupling, it is necessary to break the spin symmetry.   
    Here, we consider two symmetry-breaking scenarios that rely on the spin-rotation interaction in $^2\Sigma$ molecules  and on mw-dressed rotational states of $^1\Sigma$ molecules.  
    
    As a specific example, consider  the YO($^2\Sigma$) molecule recently laser cooled \cite{Ding:20} and trapped in an optical lattice \cite{Wu:21}, see Fig.~\ref{fig:XXZ_YO}(e). A remarkable feature of  YO is its extremely large hyperfine interaction, which dominates over the spin-rotation interaction and results in the {total spin angular momentum} $\mathbf{G}=\mathbf{I}+\mathbf{S}$ being a good quantum number \cite{Collopy:18} at low $B$-fields, where   $\mathbf{I}$ and  $\mathbf{S}$ are the nuclear and electron spins, respectively. In the large $B$-field limit, the spin-rotational states of YO($^2\Sigma$)   $\ket{\tilde{N}M_N,M_S M_I}$, have well-defined values of  nuclear ($M_I$) and electron  ($M_S$) spin projections on the field axis.

\begin{figure}[t]
\includegraphics[scale=0.4,trim = 20 0 0  0]{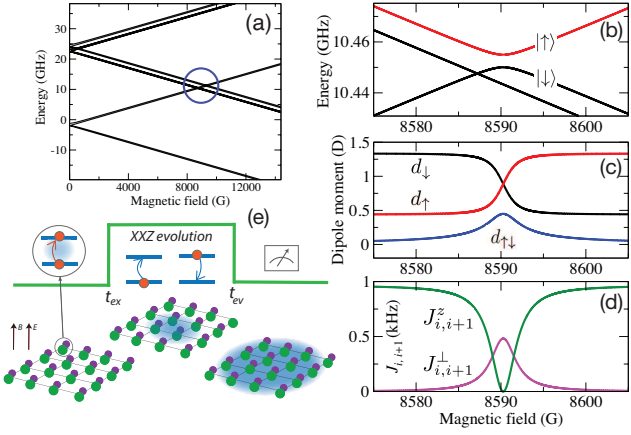}
\caption{(a) Zeeman energy levels of YO at $E=5$ kV/cm.   The ALC between the levels  $\ket{\tilde{0}0,\frac{1}{2}M_I}$ and  $\ket{\tilde{1}1,{-\frac{1}{2}}M_I}$  is marked with a circle.   Energy levels (b), EDM matrix elements (c), and NN spin coupling constants $J^z_{i,i+1}$ and $J^\perp_{i,i+1}$ (d)  plotted as a function of $B$-field near the ALC at  the {NN spacing} of 500~nm. (e) Experimental protocol for creating long-lived spin-squeezed states of YO molecules.}
\label{fig:XXZ_YO}
\end{figure}

    Figure~\ref{fig:XXZ_YO}(a) shows the lowest rotational energy levels of YO$(^2\Sigma)$. The opposite-parity levels $\ket{\tilde{0}0,\frac{1}{2}M_I}$ and  $\ket{\tilde{1}1,{-\frac{1}{2}}M_I}$
      cross at $B_c\simeq 0.85$~T  \cite{Friedrich:00,Tscherbul:06,Abrahamsson:07}.
At $B=B_c$ and $E>0$, the {electron} spin-rotation interaction mixes the   levels, leading to an avoided level crossing (ALC) shown in Fig.~\ref{fig:XXZ_YO}(b), {and we choose} our effective  spin-1/2 states $\ket{\alpha}=\ket{\uparrow},\ket{\downarrow}$ as 
 \begin{equation}\label{qubit_states_res}
\ket{\alpha} =  c_{\alpha 1}\ket{\tilde{0}0,\frac{1}{2}M_I} + c_{\alpha 2}\ket{\tilde{1}1,{-\frac{1}{2}}M_I},
\end{equation}
{where $c_{\alpha i}$ are $B$-dependent mixing amplitudes  \cite{SM}.}
Thus, in the vicinity of the ALC, the electron spin and rotation mixing provides $d_{\uparrow\downarrow}\ne 0$, which gives rise to the XXZ interaction   \cite{SM}
 \begin{equation}\label{XXZ}
\hat{H}_{\text{dip}}^\text{ex}=\sum_{i> j}
\Bigl{[} \frac{1}{2}J^\perp_{ij}(\hat{S}_i^+\hat{S}_{j}^- + \text{H.c.}) + J^z_{ij} \hat{S}_i^z\hat{S}_{j}^z \Bigr{]},
\end{equation}
where $J^\perp_{ij}=\frac{1{-}3\cos^2\theta_{ij}}{|\mathbf{R}_{ij}|^3}  2d_{\uparrow\downarrow}^2$ is the spin-exchange coupling constant.
As shown  in Figs.~\ref{fig:XXZ_YO}(c)-(d), because the spin mixing in Eq.~(\ref{qubit_states_res}) is localized in the vicinity of the ALC, both $d_{\uparrow\downarrow}$ and $J^\perp_{ij}$ peak at $B=B_c$, where $|c_{\alpha i}|=1/\sqrt{2}$.   The diagonal matrix elements of the EDM become equal at $B=B_c$, where $J^z_{ij}$ vanishes [see Fig.~\ref{fig:XXZ_YO}(d)].
The remarkable magnetic tunability of the spin coupling constants near ALCs can be used to vary the ratio $J^z_{ij}/J^\perp_{ij}$ over a wide range.
 Achieving such tuning of  $J^z_{ij}/J^\perp_{ij}$ is desirable for accessing new regimes of spin-squeezing dynamics depending on the sign of $J_z - J_\perp$ \cite{Perlin:20} and quantum simulation, since its tunability is far from straightforward in traditional  protocols that rely on pure rotational states, requiring the use of multiple mw fields \cite{Gorshkov:11b}.

  To realize the spin-squeezed states experimentally, we propose to use a standard Ramsey protocol shown in Fig.~4(e), in which mw fields are used to  excite the initial coherent superposition{$\frac{1}{\sqrt{2}}[\ket{\uparrow} + \ket{\downarrow}]$ (as recently realized in a beam experiment  \cite{Altuntas:18})}  and during the dark time the system is allowed to evolve under the XXZ Hamiltonian. 
 
The evolution leads to the formation of a spin-squeezed state in the $\{\ket{\uparrow},\ket{\downarrow}\}$ basis \cite{Bilitewski:21}. At the last step, the $B$-field is adiabatically ramped down to transfer the $\ket{\uparrow}$ and $\ket{\downarrow}$ states back to the $|\tilde{0}0\frac{1}{2} M_I\rangle$ and $|\tilde{1}1-\frac{1}{2} M_I\rangle$ zeroth-order states, and the latter state is coherently transferred to the $\ket{\tilde{0}0-\frac{1}{2}M_I}$ state using a circularly polarized mw $\pi$-pulse. As a result, we obtain a long-lived squeezed state of electron spins of $\tilde{N}=0$ $^2\Sigma$ molecules, which can be used to measure, e.g., external $B$-fields.
{The effects of decoherence on spin-squeezed states can be efficiently mitigated \cite{SM} as long as the evolution  time remains much smaller than the single-molecule coherence time, as is currently the case for, e.g.,  YO and CaF.}


 \begin{figure}[t]
\includegraphics[scale=0.51,trim = 0 0 0  0]{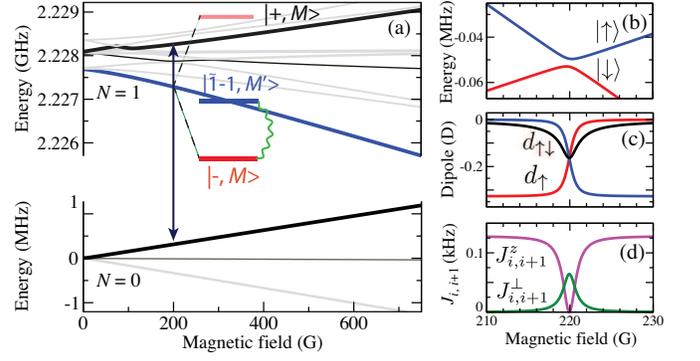}
\caption{(a)  Mw-dressed states $|\pm,M\rangle$ (red bars) of $^{40}$K$^{87}$Rb for $M_F=7/2$ obtained by driving the mw transition (vertical arrow, $\Omega/2\pi = 2.1$~MHz) between the bare levels $\ket{\tilde{0}0,-2,-\frac{3}{2}}$ and $\ket{\tilde{1}0,-2,-\frac{3}{2}}$. The effective spin-1/2 is encoded into the  mw-dressed state $|-,M\rangle$ and the bare state $\ket{\tilde{1} -1, -4, \frac{3}{2}}$ (blue bar), which are coupled by the electric quadrupole interaction (wavy line).  $B$-field dependence of the effective spin-1/2 energy levels (b),  EDM matrix elements (c), and  NN XXZ spin coupling constants $J^\perp_{i,i+1}$  and $J^z_{i,i+1}$ (d) near the ALC.}
\label{fig:XXZ_KRb}
\end{figure}


The above scenario of generating magnetically tunable  XXZ interactions
is clearly unfeasible for $^1\Sigma$ molecules due to their extremely small nuclear magnetic moments. To overcome this limitation, we use mw dressing to make the opposite-parity states degenerate as illustrated in Fig.~\ref{fig:XXZ_KRb}(a). 
Specifically,  driving the transition between the states $\ket{\tilde{0}0,M} $ and $ \ket{\tilde{1}0,M}$ (such as the $\ket{\tilde{0}0,-2,-\frac{3}{2}}$ and  $\ket{\tilde{1}0,-2,-\frac{3}{2}}$ states  of KRb)  by a linearly polarized mw field  creates a pair of mw-dressed states in the rotating frame \cite{Gorshkov:11a}
\begin{equation}\label{mw_dressed_states}
\ket{\pm,M} = c_{\tilde{0}}^\pm (\Omega,\Delta) \ket{\tilde{0}0,M} + c_{\tilde{1}}^{\pm}(\Omega,\Delta) \ket{\tilde{1}0,M},
\end{equation}
where the mixing coefficients $c_{\tilde{N}}^\pm$ depend on the Rabi frequency  $\Omega$ and the detuning from  resonance  $\Delta$.  For simplicity, we will consider the case of resonant driving ($\Delta=0$), where the energies of the mw-dressed states  $E_{\pm}=E_{1,M}\pm \Omega/2$ (assuming $\hbar=1$), and $E_{1,M}$ is the energy of the bare state $\ket{\tilde{1}0,M}$.
We note that the  coherence time of these states could be limited by the different trapping potentials experienced by the $\ket{\tilde{0}0,M}$ and $\ket{\tilde{1}0,M}$ bare states, as well as by the coherence properties of the dressing field. 
Fortunately, it may be possible to  choose the mixing coefficients $c_{\tilde{N}}^\pm$ in such a way as to achieve state-insensitive trapping conditions \cite{Gorshkov:11b,Tscherbul:22b}.

To encode the effective spin-1/2, we choose the lowest-energy mw-dressed state $\ket{-,M}$ defined above and the bare state $\ket{\tilde{1}-1,M'}=\ket{\tilde{1} -1, -4, \frac{3}{2}}$ of KRb shown in Fig.~\ref{fig:XXZ_KRb}(a).
These states are  coupled by the electric quadrupole interaction \cite{SM} via the matrix element $V_{MM'} = \bra{-,M} \hat{H}_{eQ} \ket{\tilde{1}-1,M'}=1.8$~kHz.
Figure~\ref{fig:XXZ_KRb}(b) shows the energies of  {our effective spin-1/2 states} $\ket{\uparrow,\downarrow}=c_M^{(\uparrow\downarrow)}\ket{-,M}+c_{M'}^{(\uparrow\downarrow)}|\tilde{1}-1,M'\rangle$  
as a function of $B$-field.
The NN spin coupling constants $J^\perp_{i,i+1}$ and $J^z_{i,i+1}$ shown in Fig.~\ref{fig:XXZ_KRb}(d) reach their maximal and  minimal values at $B=B_c$. We note that  $J^z_{ij}(B_c)=0$ since $d_{\uparrow}=d_{\downarrow}$ at the ALC, and thus the ratio $J^\perp_{i,i+1}/J^z_{i,i+1}$ can be magnetically tuned over a wide dynamic range.


In summary, we have shown how to engineer long-lived cluster and spin-squeezed states using the nuclear spins of $^1\Sigma$ molecules and the electron spins of $^2\Sigma$ molecules in their ground rotational states. 
The proposed schemes can be applied to a wide range of polar  molecules recently cooled and trapped in many laboratories 
\cite{Ospelkaus:10a,Caldwell:20,Burchesky:21,Collopy:15,Collopy:18,Ding:20,Wu:21,Truppe:17,Anderegg:18,Barry:14,McCarron:18},
opening up a general path to long-lived spin entanglement generation in ultracold molecular ensembles.

\begin{acknowledgments}
We thank Junru Li and Jeremy Young for careful reading
and feedback on the manuscript, and Rebekah Hermsmeier for verifying some of our results. This work was supported
by the NSF EPSCoR RII Track-4 Fellowship (Grant
No. 1929190), the AFOSR MURI, the NSF JILA-PFC
PHY-1734006, and by NIST.
\end{acknowledgments}

\widetext
\begin{center}
\textbf{\large Supplemental Material}
\end{center}
\setcounter{equation}{0}
\setcounter{figure}{0}
\setcounter{table}{0}
\setcounter{page}{1}

\author{Timur V. Tscherbul$^{1}$ Jun Ye$^{2}$, and Ana Maria Rey$^{2}$}

\affiliation{$^{1}$Department of Physics, University of Nevada, Reno, Nevada, 89557, USA}
\affiliation{$^{2}$JILA, National Institute of Standards and Technology, and Department of Physics, University of Colorado,  Boulder, Colorado, 80309, USA}
\maketitle

In this Supplemental Material, we present technical details concerning molecular Hamiltonians and  the computation of the energy levels  of KRb and YO molecules in the presence of external electric and magnetic fields, {along with a detailed analysis of decoherence effects on cluster and spin-squeezed states and of the experimental feasibility of our protocols}.  We also provide a derivation of the interaction Hamiltonian between polar $^2\Sigma$ molecules near avoided level crossings (ALCs) of their opposite-parity Zeeman states, and show that it takes the form of the XXZ spin model in the immediate vicinity of an ALC, and of the quantum Ising model far away from the ALC.

\section{Hamiltonians and energy levels of $^1\Sigma$ and $^2\Sigma$ molecules in electric and magnetic fields}

\subsection{$^1\Sigma$ molecules (KRb)}

To calculate the energy levels of  KRb($^1\Sigma^+$) as a function of external dc  electric and magnetic fields, we use the effective Hamiltonian for the ground vibrational state \cite{Aldegunde:08}
\begin{equation}\label{Hmol_KRb}
\hat{H}_\text{mol}= B_e \hat{N}^2 + \hat{H}_\text{hf} + \hat{H}_{E} + \hat{H}_{B},
\end{equation}
where the first term describes the rotational energy of the molecule, $B_e$ is the rotational constant, and $\mathbf{\hat{N}}$ is the rotational angular momentum operator.  The second term in Eq.~(\ref{Hmol_KRb}) represents the hyperfine interaction, and the third and fourth terms describe the interaction of the molecule with external dc electric and magnetic fields.

The hyperfine Hamiltonian is given by
\begin{equation}\label{Hhf}
\hat{H}_\text{hf} =  \hat{H}_{eQ} + \hat{H}_{IN} + \hat{H}_{ss,t} + \hat{H}_{ss,d},
\end{equation}
where the first term on the right-hand side stands for the electric quadrupole interaction, the second term for the  nuclear spin-rotation interaction, the third term for the scalar spin-spin interaction, and the fourth term for the tensor spin-spin interaction. 
The electric quadrupole interaction provides the dominant contribution to the hyperfine coupling in KRb ($\simeq 1$ MHz), followed by the nuclear spin-rotation and scalar spin-spin couplings ($10$s of kHz) and by the tensor spin-spin coupling ($\simeq 10$ Hz).

The Zeeman interaction of the molecule with an external magnetic field is given by 
\begin{equation}\label{H_B}
\hat{H}_{B} = -g_r\mu_N BN_z - g_1\mu_N BI_{1_z}(1-\sigma_1) - g_2\mu_NBI_{2_z}(1-\sigma_2)
\end{equation}
where $I_{\nu_z}$ are the nuclear spin angular momentum operators along the $z$-axis defined by the external magnetic field vector $\mathbf{B}$, the subscripts $\nu=1,2$ refer to the individual nuclei (K and Rb),   $B=|\mathbf{B}|$,  $g_r$ is the rotational $g$-factor, $g_i$ are the nuclear $g$-factors,  $\sigma_i$ are the diagonal elements of the  nuclear shielding tensor \cite{Aldegunde:08}, and $\mu_N$ is the nuclear Bohr magneton.

Finally, the interaction of the molecule with an external electric field is described by
\begin{equation}\label{H_E}
\hat{H}_{E}  = -\mathbf{E} \cdot \mathbf{d} = -E{d} \cos\theta,
\end{equation}
where $d$ is the magnitude of the electric dipole moment vector $\mathbf{d}$, $\mathbf{E}$ is the electric field vector, and $\theta$ is the angle between the molecular axis and the quantization axis defined by $\mathbf{E}$. We assume that the electric and magnetic field vectors are parallel ($\mathbf{E}\, ||\, \mathbf{B}$).

The molecular constants that parametrize the Hamiltonian (\ref{Hmol_KRb}) are taken from Ref.~\cite{Aldegunde:08} for the fermionic $^{40}\text{K}^{87}\text{Rb}$ isotope. To compute the energy levels we constructed and diagonalized the Hamiltonian  in the uncoupled basis $\ket{NM_N}\ket{I_1 M_{I_1}} \ket{I_2 M_{I_2}}$ using the expressions for the matrix elements from Ref.~\cite{Gorshkov:11b}. A total of 5 rotational basis states are included in the basis to produce converged energy levels at electric fields up to 30 kV/cm. To verify our computed energy levels, we compared them to  the data shown in Fig.~1 of Ref.~\cite{Aldegunde:09} and found good agreement.

\subsection{$^2\Sigma$ molecules (YO)}

The Hamiltonian of  a $^2\Sigma$ molecule such as YO$(^2\Sigma+)$ is given by \cite{Collopy:15}
\begin{equation}\label{Hmol_YO}
\hat{H}_\text{mol}= B_e \hat{N}^2 + \gamma \mathbf{\hat{N}}\cdot \mathbf{\hat{S}} + \hat{H}_\text{hf} + \hat{H}_{E} + \hat{H}_{B},
\end{equation}
where the first term represents the rotational energy, and the second term describes the spin-rotation interaction of $\mathbf{N}$ with $\mathbf{S}$, the electron spin angular momentum of the molecule, whose strength is quantified by the spin-rotation constant $\gamma$.  The hyperfine interaction is given by the third term in Eq.~(\ref{Hmol_YO}), whereas the interaction of the molecule with external dc electric and magnetic fields is described by  the third and fourth terms. The hyperfine interaction has the form
\begin{equation}\label{Hhf_YO}
\hat{H}_\text{hf}= (b+c/3)\mathbf{I}\cdot \mathbf{S} + c\frac{\sqrt{6}}{3} \left(\frac{4\pi}{5}\right)^{1/2} \sum_q (-1)^q Y_{2,-q}(\mathbf{r})[\mathbf{I}\otimes \mathbf{S}]^{(2)}_q,
\end{equation}
where $b$ and $c$ are the isotropic and anisotropic hyperfine constants, respectively, $\mathbf{r}$  describes the orientation of the molecular axis in the laboratory frame with the $z$-axis defined by the external  magnetic fields (see above), $ Y_{2,-q}$ is a spherical harmonic, and $[\mathbf{I}\otimes \mathbf{S}]^{(2)}_q$ is a second-rank tensor product of the electron and nuclear spin operators \cite{Brown:03}.
The hyperfine interaction in YO remarkably strong, making the total spin $|\mathbf{G}|=|\mathbf{I}+\mathbf{S}|$ a good quantum number in the weak $B$-field limit. The interaction with an external magnetic field $\hat{H}_B=g_S \mu_0 BS_z$, where $\mu_0$ is the Bohr magneton and $g_S\simeq 2$ is the electron spin $g$-factor.

The molecular constants $B_e$, $\gamma$, $b$, and $c$ for YO are taken from Ref.~\cite{Collopy:15} and the Hamiltonian (\ref{Hmol_KRb}) is represented in the basis $\ket{NM_N}\ket{SM_S}\ket{IM_I}$ composed of eigenstates of $\mathbf{N}^2$ and $\hat{N}_z$, $\mathbf{S}^2$ and $\hat{S}_z$, and $\mathbf{I}^2$ and $\hat{I}_z$.

The matrix elements of the operators in Eq. (\ref{Hmol_YO}) are evaluated as described in  Ref.~\cite{Tscherbul:07}. Numerical diagonalization of the Hamiltonian matrix provides the energy levels of YO  in good agreement with the literature data from Ref.~\cite{Collopy:15}.

\section{Derivation of single-molecule effective Hamiltonian and of the spin lattice Hamiltonian for $^2\Sigma$ molecules near avoided crossings}

\subsection{Effective  Hamiltonian for a single $^2\Sigma$ molecule in its ground rovibrational state near an ALC}

Consider a single $^2\Sigma$ 
molecule in superimposed electric and magnetic fields described by the Hamiltonian \cite{Friedrich:00,Tscherbul:06,Abrahamsson:07}
\begin{equation}\label{Hmol}
\hat{H}_\text{mol} = B_e \hat{N}^2 + \gamma \hat{\mathbf{N}} \cdot \hat{\mathbf{S}} - g_S\mu_0 B \hat{S}_z  - \hat{\mathbf{d}}\cdot \mathbf{E}
\end{equation}
where  $\hat{\mathbf{N}}$ and $\hat{\mathbf{S}}$  are the rotational and spin angular momentum operators of the molecule, $B_e$ is the rotational constant, $\gamma$ is the spin-rotation constant, $g_S\simeq 2.0$ is the  $g$-factor, and $\mu_0$ is the Bohr magneton.  The quantization axis in Eq.~(\ref{Hmol}) is chosen along the direction of the external magnetic field, which is assumed to be parallel to the electric field. We will neglect the molecular hyperfine structure, which is an excellent approximation at large magnetic fields required to induce avoided crossings between the different rotational levels in $^2\Sigma$ molecules. For example, in YO($^2\Sigma^+$), which has an extremely large hyperfine interaction,  the level crossings occur near 8500 G [see Figs.~4(a) and 4(b) of the main text], where  the nuclear spin is completely decoupled from the electron spin and from molecular rotation, and thus $M_I$ remains a good quantum number. Because  the molecular  Hamiltonian  does not couple  the states $\ket{\tilde{0}0\frac{1}{2}M_I}$ and $\ket{\tilde{1}1-\frac{1}{2}M_I'}$  involved in the ALC unless $M_I = M_I'$, the ALCs can only occur between the  states of the same $M_I$.

 We will use the basis $|NM_N\rangle|SM_S\rangle=\ket{NM_NM_S}$, where $|NM_N\rangle$ are the eigenstates of $\hat{N}^2$ and $\hat{N}_z$ and  $|SM_S\rangle$ are those of $\hat{S}^2$ and $\hat{S}_z$ [we will omit the label $S$ since $S=1/2$ for $^2\Sigma^+$ molecules]. Near  the ALC between the opposite-parity Zeeman levels [see Fig.~4(b) of the main text], we only need to take  into account  three  basis states in the weak $E$-field limit: $\ket{00\frac{1}{2}}$, $\ket{10\frac{1}{2}}$, and $\ket{11-\frac{1}{2}}$. In this basis, the matrix of the molecular Hamiltonian (\ref{Hmol}) takes the form \cite{Abrahamsson:07}
\begin{equation}\label{Hmol_mat}
\mathsf{H}_\text{mol} = \begin{pmatrix}
 \mu_0 B &  -{Ed}/{\sqrt{3}} & 0 \\
-{Ed}/{\sqrt{3}} & 2B_e + \mu_0 B &  \gamma/\sqrt{2}  \\
0 &  \gamma/\sqrt{2}  & 2B_e - \mu_0 B - \gamma/2 &  \\
  \end{pmatrix}.
  \end{equation}
Here, the basis states $\ket{00\frac{1}{2}}$ and $\ket{10\frac{1}{2}}$ are coupled by an external electric field 
and the basis states  $\ket{10\frac{1}{2}}$, and $\ket{11-\frac{1}{2}}$ are coupled by the spin-rotation interaction. 

In the weak electric field limit, where $\beta_E={Ed}/B_e \ll 1$, we can define the zeroth-order  Zeeman states, which do not interact except in close vicinity of the ALC:
\begin{align}\label{ALCstates}\notag
&\ket{\tilde{0}0\frac{1}{2}} = c_0 \ket{00\frac{1}{2}} + c_1 \ket{10\frac{1}{2}},   \\
&\ket{\tilde{1}1-\frac{1}{2}} = \ket{11-\frac{1}{2}}.
\end{align}
where we have neglected the electric field coupling between the $N=1$ and $N=2$ rotational states, since it does not affect the interaction between the states (\ref{ALCstates}) to first order in $\beta_E$.

Projecting the molecular Hamiltonian (\ref{Hmol}) onto the zeroth-order basis (\ref{ALCstates}) and using Eq. (\ref{Hmol_mat}) we obtain the matrix elements
\begin{align}\label{Hmol2x2deriv}\notag
\langle \tilde{0}0\frac{1}{2} | \hat{H}_\text{mol} | \tilde{0}0\frac{1}{2}\rangle &= c_0^2 \mu_0B + 2 c_0 c_1 (-Ed/\sqrt{3}) + c_1^2 (2B_e+\mu_0 B),  \\ \notag
\langle \tilde{0}0\frac{1}{2} | \hat{H}_\text{mol} | \tilde{1}1-\frac{1}{2}\rangle &= c_1 \langle 10\frac{1}{2}|\hat{H}_\text{mol}|11-\frac{1}{2}\rangle = c_1 \gamma/\sqrt{2},   \\
\langle \tilde{1}1-\frac{1}{2} | \hat{H}_\text{mol} | \tilde{1}1-\frac{1}{2}\rangle &=  \langle 11-\frac{1}{2}|\hat{H}_\text{mol}|11-\frac{1}{2}\rangle = 2B_e-\mu_0 B - \gamma/2.   
\end{align}
These equations can be simplified in the weak $E$-field limit, where $c_0 \to 1$ and $c_1\to Ed/(2\sqrt{3}B_e)$ \cite{Wall:15c}:
\begin{align}\label{Hmol2x2}\notag
\langle \tilde{0}0\frac{1}{2} | \hat{H}_\text{mol} | \tilde{0}0\frac{1}{2}\rangle &= \mu_0B - \frac{(Ed)^2}{6B_e}, \\ \notag
\langle \tilde{0}0\frac{1}{2} | \hat{H}_\text{mol} | \tilde{1}1-\frac{1}{2}\rangle &= \frac{Ed \gamma}{2\sqrt{6}B_e},   \\
\langle \tilde{1}1-\frac{1}{2} | \hat{H}_\text{mol} | \tilde{1}1-\frac{1}{2}\rangle &=   2B_e-\mu_0 B - \gamma/2.   
\end{align}

Thus, the energy level structure of a $^2\Sigma$ molecule near the lowest-energy ALC of its opposite-parity Zeeman levels is described by the following effective Hamiltonian in the zeroth-order basis $\{ \ket{\tilde{0}0\frac{1}{2}}, \ket{\tilde{1}1-\frac{1}{2}} \}$
\begin{equation}\label{Heff_mel}
\mathsf{H}_\text{eff} = \begin{pmatrix}
 \mu_0 B - (Ed)^2/(6B_e) &  {Ed\gamma}/(2\sqrt{6} B_e) \\
  {Ed\gamma}/(2\sqrt{6} B_e) & 2B_e - \mu_0 B -  \gamma/2  \\
  \end{pmatrix}
  \end{equation}
  The diagonal matrix elements of the effective Hamiltonian give the energies of zeroth-order Zeeman levels, which become equal at the crossing point defined by $2\mu_0 B_c=2B_e-\gamma/2+ (Ed)^2/(6B_e)$. The off-diagonal matrix element  $Ed\gamma/(2\sqrt{6}B_e)$ quantifies the coupling between the  zeroth-order Zeeman  levels,
which requires both the electric field (to couple the spatial components of bare $N=0$ and $N=1$ rotational states) {\it and} the spin-rotation interaction (to couple the spin components of the $N=1$ rotational states).
  
  The eigenstates near the ALC,   $\ket{\alpha}=\{ \ket{\uparrow},\ket{\downarrow} \}$,
   can be obtained by diagonalizing the effective Hamiltonian (\ref{Heff_mel}) 
  \begin{equation}\label{expansion}
\ket{\alpha} = {c_{\alpha \tilde{0}}} \ket{\tilde{0}0\frac{1}{2}} + c_{\alpha \tilde{1}} \ket{\tilde{1}1-\frac{1}{2}},
\end{equation}
where $(c_{\alpha \tilde{0}}, c_{\alpha \tilde{1}})^T$  are the eigenvectors of the $2\times 2$ effective Hamiltonian matrix (\ref{Heff_mel}) corresponding to the eigenvalues $E_\alpha$.
At the crossing point ($B=B_c$) the diagonal matrix elements 
are degenerate, and thus even an arbitrarily small electric field results in a complete mixing of the opposite-parity states ($|c_{\alpha i}| = 1/\sqrt{2}$).  
 {\it  The  effective spin-1/2  states near the ALC are  thus linear combinations of the zeroth-order Zeeman states of the opposite parity and spin}.

\subsection{Effective spin-spin interactions and spin lattice Hamiltonian for  $^2\Sigma$ molecules near avoided level crossings}

In this section, we derive the electric dipolar (ED)  interaction Hamiltonian and the many-body spin Hamiltonian for polar $^2\Sigma$ molecules in the vicinity of an ALC. We begin by considering two $^2\Sigma$ molecules $i$ and $j$ with the dipole moment operators $\hat{d}_i$ and $\hat{d}_j$ interacting via the ED interaction.
Assuming that the interaction is much smaller than the energy gap between our effective spin-1/2 states at the ALC (true in our case), we can restrict attention to energy-conserving resonant exchange terms, and consider only the $q=0$ tensor component of the ED interaction \cite{Wall:15c}:
\begin{equation}\label{Hdip_q0}
\hat{H}_{\text{dip},ij}^{(q=0)} = \frac{1{-}3\cos^2\theta_{ij}}{R_{ij}^3} \Bigl{[} \frac{1}{2}(\hat{d}_i^{(1)} \hat{d}_{j}^{(-1)} + \hat{d}_i^{(-1)} \hat{d}_{j}^{(1)}) \\ 
 + \hat{d}_i^{(0)} \hat{d}_{j}^{(0)}  \Bigr{]},
\end{equation}
where
 $\mathbf{R}_{ij}$ joins the centers of mass of the molecules, $R_{ij}=|\mathbf{R}_{ij}|$, and $\theta_{ij}$ is the angle between $\mathbf{R}_{ij}$ and the quantization axis defined by the external dc electric and magnetic fields.

To derive an effective spin-spin interaction Hamiltonian between two $^2\Sigma$ molecules in the effective spin-1/2 states $\ket{\alpha} = \ket{\uparrow}, \ket{\downarrow}$  given by Eq. (\ref{expansion}), we project the dipole-dipole interaction (\ref{Hdip_q0}) onto the two-qubit basis $\ket{\alpha_i\alpha_j} = \{ \ket{\uparrow_i\uparrow_j}, \ket{\uparrow_i\downarrow_j}, \ket{\downarrow_i\uparrow_j}, \ket{\downarrow_i\downarrow_j} \}$  to obtain
\begin{equation}
\hat{H}_{\text{dip},ij}^{(q=0)} = \frac{1{-}3\cos^2\theta_{ij}}{R_{ij}^3} \sum_{ \alpha_i,\alpha_j} \sum_{\alpha_i',\alpha_j'}  
 \bra{\alpha_i} d^{(0)}_i \ket{\alpha_i'} 
 \bra{\alpha_j} d^{(0)}_j \ket{\alpha_j'} \ket{ \alpha_i\alpha_j} \bra{ \alpha_i'\alpha_j'}
\end{equation}
Evaluating the matrix elements using Eq. (\ref{expansion}) and the fact that the operators $\hat{d}^{(0)}_i$ are diagonal in the electron spin quantum number $M_S$, we arrive at a $4\times 4$ matrix representation of the ED Hamiltonian 
\begin{equation}\label{Hex_matrix}
\mathsf{H}_{\text{dip},ij}=\frac{1{-}3\cos^2\theta_{ij}}{R_{ij}^3}
\begin{pmatrix}
d_{\uparrow}^2 & 0 & 0 & 0 \\
0 & d_{\uparrow}d_{\downarrow} & d_{\downarrow\uparrow}^2  & 0  \\
0 & d_{\downarrow\uparrow}^2 & d_{\uparrow}d_{\downarrow}  & 0  \\
0 & 0 & 0 & d_{\downarrow}^2
\end{pmatrix},
\end{equation}

Introducing the effective spin-1/2 operators $\hat{S}_i^z$, $\hat{S}_i^+$, and $\hat{S}_i^-$ acting in the two-dimensional Hilbert subspace of the $i$-th molecule $\{{\ket{\downarrow},\ket{\uparrow}} \}$, the ED interaction (\ref{Hex_matrix}) can be rewritten as a spin-spin interaction Hamiltonian \cite{Wall:15c}
\begin{equation}\label{Hij_ex_spin}
\hat{H}_{\text{dip},ij}=
\frac{1{-}3\cos^2\theta_{ij}}{R_{ij}^3} \Bigl{[} \frac{J_\perp}{2} (\hat{S}_i^+\hat{S}_{j}^- + \text{H.c.})   + J_z \hat{S}_i^z\hat{S}_{j}^z  + W_z(\hat{1}_i \hat{S}_{j}^z + \hat{S}_{i}^z \hat{1}_j ) + V \hat{1}_i\hat{1}_j \Bigr{]},
\end{equation}
where $\hat{1}_i$ is the unit operator in the effective two-state space of the $i$-th molecule, and the  spin coupling constants are expressed in terms of the dipole matrix elements \cite{Wall:15c}
\begin{align}\label{Hij_ex_spin_constants}\notag
J_\perp &= 2d_{\downarrow\uparrow}^2, \quad  J_{z}= (d_{\uparrow} - d_{\downarrow})^2, \\
W_z &= \textstyle{\frac{1}{2}} (d_{\uparrow}^2 - d_{\downarrow}^2),\quad V = \textstyle{\frac{1}{4}} (d_{\uparrow} + d_{\downarrow})^2
\end{align}

To derive the many-body lattice spin Hamiltonian for interacting $^2\Sigma$ molecules near an ALC, we  sum the pairwise interactions given by Eq.~(\ref{Hij_ex_spin}) over all lattice sites  $\hat{H}_\text{dip}=\sum_{i,j}\hat{H}_{\text{dip},ij}$ \cite{Wall:15c}.
 We further replace the identity operators $\hat{1}_i$ with the total molecular density operator $\hat{n}_i$, and  assume unit filling and homogeneous density of molecules in the lattice, which allows us to neglect the terms proportional to $W_z$ and $V$ in Eq.~(\ref{Hij_ex_spin}) as constant energy offsets. The resulting spin Hamiltonian for pinned $^2\Sigma$ molecules in a unit-filled lattice takes the form of the XXZ Hamiltonian [see Eq.~(3)  of the main text]
 \begin{equation}\label{H_spin_mb}
\hat{H}_{\text{dip}} = \sum_{i>j} 
\Bigl{[} \frac{J_{ij}^\perp}{2} (\hat{S}_i^+\hat{S}_{j}^- + \text{H.c.})   + J_{ij}^z \hat{S}_i^z\hat{S}_{j}^z \Bigr{]},
\end{equation}
where $J_{ij}^\beta =\frac{1{-}3\cos^2\theta_{ij}}{R_{ij}^3} J_\beta$ ($\beta=||,\perp$). 

Far away from the ALC, the spin exchange coupling constant $J_{ij}^\perp$ approaches zero as shown in Fig. 4(d) of the main text, and Eq.~(\ref{H_spin_mb}) reduces to the Ising Hamiltonian [Eq.~(1) of the main text]
\begin{equation}\label{H_Ising}
\hat{H}_{\text{dip}} = \sum_{i>j}
 J_{ij}^z \hat{S}_i^z\hat{S}_{j}^z.
\end{equation}

\section{Detailed analysis of decoherence effects}

{Assuming that two-body losses due to molecule-molecule collisions are slow, which is a good approximation for the lattice configuration proposed in this work, where molecules are strongly localized in their individual lattice sites,  relaxation $(T_1)$ processes can be safely neglected}. The main source of decoherence for lattice-confined molecules is then pure dephasing ($T_2$) noise due to the different trapping potentials experienced by the  $\tilde{N}=0$ and $\tilde{N}=1$ rotational states \cite{Kotochigova:10b,Neyenhuis:12,Yan:13,Tobias:22} because of their different polarizabilities and  Gaussian profiles of the laser beams generating the lattice. 
 This noise currently limits spin-echo coherence times of  rotational (and likely spin-rotational) superposition states to 470 ms \cite{Burchesky:21}.

A distinct advantage of our protocols is that  most of the time, the molecules  remain in the superpositions of the different nuclear spin sublevels of a single rotational state, which have extremely long coherence times. Thus, significant decoherence can only occur during dynamical evolution under the Ising or XXZ Hamiltonians, when the molecules are in superpositions of the spin-rotational states  $\ket{\tilde{0}0,M}$ and $\ket{\tilde{1}0,M'}$ (see Figs.~2(c) and 4(e) of the main text).
Below we will consider the effects of decoherence on dynamical evolution under the Ising Hamiltonian, which creates cluster states, and under the XXZ  Hamiltonian, which produces spin-squeezed states. 

\vskip9pt
{\bf Effect of decoherence on cluster states} 
\vskip3pt

We first consider the effect of decoherence on cluster states generated by time evolution under the Ising Hamiltonian. 
The Ising Hamiltonian commutes with the pure dephasing noise (we assume that $\hbar=1$)
\begin{equation}\label{Hnoise}
\hat{H}_\text{noise}= \sum_{i} h_i(t)\hat{S}^z_i,
\end{equation}
 where $h(i)$ are random variables representing zero-mean uncorrelated Gaussian noise with the autocorrelation function $\langle h_i(t) h_j(t')\rangle=\delta_{ij}f(t-t')$. The Gaussian noise satisfies $\langle \exp [-i\int_0^t h(\tau)d\tau]\rangle = \exp[-\Gamma(t)]$, where $\Gamma(t) = \frac{1}{2} \int_0^t dt_1 \int_0^t dt_2 f(t_1-t_2)$. 
Because the static dephasing noise ($h_i(t)=h_i$) can be eliminated using  spin echo techniques \cite{Yan:13}, the relevant timescale  is the spin echo rotational coherence time $T_2=470$ ms, which is much longer than our estimated cluster time of 2.95~ms (see main text). Thus, we expect the effects of decoherence during the dynamical Ising evolution to be minimal. 

Following previous theoretical work \cite{Rey:08} we will consider the white noise limit, where $f(t-t')=\Gamma_d\delta (t-t')$ and $\Gamma(t)=\Gamma_dt/2$, where $\Gamma_d$ is the dephasing rate.
The effects of dynamic white noise cannot be removed by spin-echo pulses, but its impact on the cluster state can be quantified by using the expectation values of stabilizer operators, which quantify cluster-state entanglement
\begin{equation}\label{stabilizer_exp}
\langle K\rangle_j = 2^{2d+1}\langle  \hat{S}_j^x \prod_{\langle j,k  \rangle} \hat{S}_k^z \rangle,
\end{equation}
where $d$ is the spatial dimension, and the average is taken over the wavefunction $\ket{\psi}$ of the many-body system. The expectation values provide a better measure than global state fidelity because it captures the localized nature of the entanglement \cite{Mamaev:19}. A perfect cluster state has $\langle K\rangle_j =\pm1$ for all lattice sites $j$, and the deviation of  $|\langle K\rangle_j|$ from unity is a measure of cluster state quality at the $j$-th site. 


To  evaluate the  time evolution of the expectation values in Eq.~\eqref{stabilizer_exp}, we follow the analytical theory developed in Ref.~\cite{Foss-Feig:13}. Because the dephasing noise commutes with the $\hat{S}^z_j$ operator on each site,  it does not contribute to the decay of the correlation function containing only $\hat{S}^z_j$ operators. On the other hand, because $\hat{S}^x_j$ does not commute with the noise, the latter causes irreversible decay of the expectation value $\langle S^x_j \rangle\propto e^{-\frac{1}{2}\Gamma_{d} t}$, where $\Gamma_d=1/T_2$ is the dephasing rate (denoted by $\Gamma_\text{el}$ in Ref.~\cite{Foss-Feig:13}). We thus obtain 
\begin{equation}\label{stabilizer_exp_dynamics}
\langle K\rangle_j (t) \simeq e^{-\frac{1}{2}\Gamma_{d} t} F(J_{jk},t),
\end{equation}
where $F(J_{jk},t)$ is a function of time and the nearest-neihbor Ising coupling constants $J_{jk}$ involving the $j$-th site. Note that $F(J_{jk},t)$ does not decay under pure dephasing noise \cite{Foss-Feig:13}, the case of interest here. 
 Thus, the decay of cluster-state quality is determined by the single-molecule dephasing rate. 

Using the spin-echo coherence time $T_2=470$~ms in Eq.~\eqref{stabilizer_exp_dynamics} 
we estimate  $|\langle K\rangle_j| = 0.9969$  at $t_c=2.95$ ms. Thus, our protocol is able to generate a high-quality ($|\langle K\rangle_j| \ge 0.95$ \cite{Mamaev:20}) nuclear spin cluster state for KRb. Molecules with larger dipole moments (and hence smaller $t_c$) will be less affected by the dephasing noise, leading to even better cluster-state quality. 



\vskip9pt
{\bf Effect of decoherence on spin-squeezed states} 
\vskip3pt

Because the dephasing noise (\ref{Hnoise}) does not commute with the XXZ Hamiltonian used to generate spin-squeezed states, it is no longer removable by the spin echo techniques, so the noise can have strongly detrimental effects on these entangled states. Below, we show how one can  mitigate these  effects using  the gap protection mechanism  (for static noise). We will also estimate the effects of the remaining dynamic  noise on the quality of spin-squeezed states.

\vskip9pt
{\it  1. Suppressing static noise using  gap protection} 
\vskip3pt

To suppress the  effects of static dephasing noise (inhomogeneous broadening)  we take advantage of a many-body gap \cite{Rey:08,Bilitewski:21}, which operates in the regime  
$|\bar{J}_\perp|> |\bar{J}_z|$ and is optimal close to the Heisenberg point {($|\bar{J}_\perp| = |\bar{J}_z$)}, where $\bar{J}_\alpha = \frac{1}{N^2}\sum_{i,j}J_{ij}^\alpha$ are the average coupling constants ($\alpha=||,\perp$), and $N$ is the number of molecules trapped in an optical lattice (in the following, we will assume a square lattice of dimension $L\times L$).

This  mechanism ensures robust  noise-protected  dynamical evolution provided the many-body energy gap  $\Delta_{MB}$ satisfies the following condition
\begin{equation}\label{gap_condition}
\Delta_{MB} > \Delta h,
\end{equation}
where $\Delta h$ is the width of the energy splitting distribution  $\Delta{E}_{\uparrow\downarrow}=E_\uparrow-E_\downarrow$ due to the different trapping potentials experienced by the $\ket{\uparrow}$ and   $\ket{\downarrow}$ effective spin-1/2 levels. 
The many-body gap for dipolar XXZ-type interactions of interest here is bounded from below as $\Delta_{MB} > \Delta_{MB}^\text{nn}$, where $\Delta_{MB}^\text{nn} = N_\text{nn} {J}^\perp_\text{nn}$ is the gap for the XXZ model with nearest-neighbor interactions,  $J^\perp_\text{nn}$ is the nearest-neighbor coupling constant, and $N_\text{nn}=4$ is  number of nearest neighbors in two dimensions.  Below we will consider a conservative scenario, where the energy gap takes its minimum value $\Delta_{MB} = 4J_\text{nn}^\perp = 4J_{i,i+1}^\perp$.

The energy splitting $\Delta{E}_{\uparrow\downarrow}$ varies from site to site, causing inhomogeneous broadening and decoherence, which are, however, suppressed by the interaction-induced many-body gap $\Delta_{MB}$ provided the latter remains large compared to the width of the energy splitting distribution \eqref{gap_condition} \cite{Rey:08}. At the lattice site with coordinates $\mathbf{R}_{ij}=a(i,j)$ the  splitting  is given by (assuming harmonic confinement)
\begin{equation}\label{Eij}
h(\mathbf{R}_{ij}) = \frac{1}{2}m (\omega_\uparrow^2 - \omega_\downarrow^2)a^2 (i^2 + j^2) = \Delta E_{\uparrow\downarrow} (i^2 + j^2),
\end{equation}
where $a$ is the lattice constant, and  $\Delta E_{\uparrow\downarrow}$ is on the order of a few Hz \cite{Yan:13}. For definiteness, we will set $\Delta E_{\uparrow\downarrow}=1$ Hz, although it can be made much smaller using box-shaped trapping potentials \cite{Gaunt:13,Mukherjee:17,Bause:21}.

The variance of the energy gap distribution on a square lattice of size $L$ ($N=L^2$) is given by
\begin{align}\label{delta_h2}\notag
(\Delta h)^2 &= \frac{1}{L^2}\left(\sum_{i,j=-L/2}^{L/2} h(\mathbf{R}_{ij})^2\right) - \bar{h}^2  \\
&=  \frac{\Delta E_{\uparrow\downarrow}^2}{L^2}\left( \sum_{i,j=-L/2}^{L/2} (i^2 + j^2)^2 -  \frac{1}{L^2}\left[ \sum_{i,j=-L/2}^{L/2} (i^2 + j^2)\right]^2  \right)
\end{align}
Evaluating the sums over $i$ and $j$ analytically and taking the limit of large $L$ gives
\begin{equation}\label{delta_h}
\Delta h \simeq \frac{L^2}{\sqrt{90}} \Delta E_{\uparrow\downarrow}.
\end{equation}
This expression gives results accurate to $\le 25$\% for $L\ge 10$.

Combining the result (\ref{delta_h}) with Eq.~\eqref{gap_condition} we obtain the maximum lattice size, for which the gap protection mechanism is effective 
\begin{equation}\label{delta_h}
L_\text{max}^2 = \frac{4J^\perp_\text{nn}\sqrt{90}}{\Delta E_{\uparrow\downarrow}}.
\end{equation}
Using $J^\perp_\text{nn}=50$~Hz and $\Delta E_{\uparrow\downarrow}=1$~Hz, we obtain $L_\text{max} = 43$ for KRb molecules. This conservative estimate (see above)  shows that the gap-protected dynamical evolution could be used to generate a robust spin-squeezed state of as many as  1849 molecules trapped in a $43\times 43$ square lattice.

Finally, we note that, in addition to the gap protection mechanism, spin-echo pulses could be applied in the gap-protected regime to further eliminate the static noise, thereby  extending the coherence time of the quantum many-body system \cite{Yan:13}.

\vskip9pt
{\it  2. Dynamical noise} 
\vskip3pt

To quantify the maximum degree of spin squeezing achievable in our system, we use the maximum gain in the phase resolution $\Delta\phi$ over that achieved with a coherent spin-polarized state \cite{Perlin:20}
\begin{equation}\label{xi_param}
\xi^{2}= \frac{(\Delta \phi)_\text{min}}{(\Delta\phi_\text{polarized})} = \min_\phi \text{var}(S_\phi^\perp)\frac{N}{|\langle \hat{\mathbf{S}}\rangle|^2},
\end{equation}
where  $\hat{\mathbf{S}}$ is the  collective spin operator with Cartesian components  $\hat{S}^\alpha=\sum_i \hat{S}_i^\alpha$ ($\alpha=x,y,z$). The value $\xi^2 <1$ signals the presence of many-body entanglement, and also quantifies the metrological gain beyond the standard quantum limit \cite{Sorensen:01,Ma:11}.

Unlike the static noise considered above, dynamical noise cannot be eliminated by either gap protection or spin-echo pulses. To estimate the effects of this noise on our spin-squeezed states, we use the theoretical result  for the  spin-squeezing parameter for the evolution under the one axis twisting Hamiltonian  and pure dephasing noise \cite{Chu:21}
\begin{equation}\label{xi2_noise}
\xi^{2} =  \frac{1+2\Gamma_dt_0}{(N\chi t_0)^2} + \frac{1}{6}N^2(\chi t_0)^4,
\end{equation}
where $\Gamma_d$ is the pure dephasing rate introduced above, $\chi$ is the coupling parameter in the effective one axis twisting (OAT) Hamiltonian, which for our case is proportional to $\bar{J}_z-\bar{J}_\perp$, and  $t_0$ is the optimal squeezing time, during which the system evolves into a maximally spin-squeezed state.  Equation (\ref{xi2_noise}) shows that in order to minimize the effects of dynamical noise on  spin squeezing, we need to ensure that $2\Gamma_d t_0\ll 1$ or $t_0\ll T_2/2$, where $T_2\simeq 470$~ms is the spin-echo coherence time (see above).

To verify whether the condition $t_0\ll T_2/2$ is experimentally realizable, we use the optimal  squeezing time for the  dipolar XXZ Hamiltonian on a 2D square lattice $t_\text{opt}$ computed using the truncated Wigner approximation \cite{Perlin:20}, which is more accurate than  the optimal squeezing time $t_0$ for the OAT Hamiltonian  (note that $t_0$ is to be regarded as a rough approximation).
Numerical calculations  \cite{Perlin:20} suggest that for $1/R^3$ interactions, operating at the XX point ($J_z=0$) is optimal for achieving the optimum amount of spin squeezing $\xi^2_\text{opt}\ll 1$ in the presence of decoherence since at this point the dynamics is not so slow (even though the gap protection mechanism  is not optimal in this regime,  see below). 
For example,   $\xi^2_\text{opt}=6.23\times 10^{-3}$ for $N=4096$ molecules in a $64\times 64$  lattice     (see the upper left panel of Fig.~2 of Ref.~\cite{Perlin:20}).

The XX point is reached at the avoided crossing in YO($^2\Sigma^+$) shown in Fig.~4(d) of the main text, where $J_z=0$, $J_\perp/2\pi=0.5$~kHz, and 
  $t_\text{opt}\simeq 4.69/J_\perp=1.5$~ms  \cite{Perlin:20}. 
   This timescale is  more than two orders of magnitude shorter than the $T_2$ time recently observed for the closely related CaF$(^2\Sigma^+)$ molecule (470 ms  \cite{Burchesky:21}). {\it  Therefore, we expect the effect of dynamic noise on the spin-squeezed states to be negligibly small.}

Importantly, as long as $T_2$ can be made significantly small compared to $t_0$, the degree of spin squeezing can be further improved by operating closer to the Heisenberg point ($J_z/J_\perp = 1$).
This has the added benefit of retaining some degree of gap protection, but comes at the cost of longer optimal squeezing times, making the system more vulnerable to decoherence. 
 This can be achieved for YO molecules by tuning the magnetic field closer to the crossing position as shown in Fig.~4(d) of the main text. For example, for $J_z/J_\perp \simeq 0.5$, we have $t_\text{opt} \simeq 10.5/J_\perp=3.4$~ms, about twice as long as the squeezing time  for $J_z/J_\perp=0$, but still much smaller than $T_2\simeq 470$ ms. For these parameters, $\xi^2_\text{opt}=5.11\times 10^{-3}$  \cite{Perlin:20}. {\it This demonstrates that it is possible to obtain a high degree of spin squeezing rapidly (compared to the $T_2$ time), while at the same time enjoying the benefits of the many-body gap protection mechanism.}

\section{Detailed analysis of experimental feasibility} 

Here, we present a list of experimental techniques  required to implement the  protocols for cluster state and spin-squeezed state generation with ultracold ensembles of KRb and YO molecules  (see Figs.~3-5 and the discussion in the main text). We will show that  most of these  techniques have already been realized experimentally.

In analyzing the effects of experimental imperfections on the quality of entangled states, we will focus on decoherence during the dynamical evolution under  the Ising or XXZ Hamiltonians.   This can be justified by recent advances in the experimental techniques for the creation of molecular arrays in optical lattices and tweezers  \cite{Kaufman:21}.
For example,  a recent experiment produced ultracold rovibrationally ground-state NaCs molecules  in the  ground   motional state of an optical tweezer \cite{Cairncross:21}.  


\vskip6pt
{\bf Experimental requirements for  cluster-state generation  with $^1\Sigma$ molecules such as KRb }

\begin{enumerate}



\item
An external dc magnetic field ($B\simeq 200{-}500$~G) to polarize the nuclear spins of alkali-dimer molecules, and a constant dc electric field ($E=0{-}20$ kV/cm) to induce electric dipolar interactions between the molecules.


The required fields have been applied in previous experiments \cite{Yan:13,Gregory:16,Gregory:21}. In particular, dc $E$-fields up to 15 kV/cm can be stably applied and temporally stabilized to 1 ppm. The dc magnetic field can be stabilized to 1 part per 1000.
Any imperfections in control fields will result in decoherence of molecular superposition states, whose effects  on entangled states are estimated in Sec.~III.

\item
A coherent superposition of different nuclear spin sublevels  $|\tilde{0}0,+\rangle$ in the ground rotational state ($\tilde{N}=0$) manifold, which forms the starting point for our cluster-state generation protocol (see the main text).

Coherent superpositions of this kind can be prepared with high fidelity using two-photon microwave transitions between the ground and first rotationally excited states, as demonstrated experimentally for KRb \cite{Ospelkaus:10a}, NaK  \cite{Park:17}, and RbCs \cite{Gregory:21}. Nuclear spin coherence times of 1~s, 5.3~s, and 3.3~s have been observed for NaK, RbCs, and NaRb, respectively \cite{Park:17,Gregory:21,Lin:22}.

\item
Single-photon microwave excitation of the  initial superposition $|\tilde{0}0,+\rangle=\frac{1}{\sqrt{2}} [ \ket{\tilde{0}0,M} +  \ket{\tilde{0}0,M'}  ]$ to produce the superposition $\ket{+}=\frac{1}{\sqrt{2}} [ \ket{\tilde{0}0,M} +  \ket{\tilde{1}0,M'}  ]$ (the initialization   
 {\it Step 1}) and deexcitation of the latter back to the   initial superposition ({\it Step 3}).  This can be achieved by applying a resonant $\pi$ pulse on the $\ket{\tilde{0}0,M'}\to \ket{\tilde{1}0,M'}$ transition, which is electric dipole-allowed since it conserves $M'$. As discussed in the main text for KRb, this transition can be energetically detuned from the competing one-photon transition $\ket{\tilde{0}0,M}\to \ket{\tilde{1}0,M}$ by applying a suitably chosen dc electric field, thus enabling selective mw driving  of the transition  $\ket{\tilde{0}0,M'}\to \ket{\tilde{1}0,M'}$.

 
Single-photon mw transitions are commonly used for highly efficient coherent state transfer in ultracold polar molecules \cite{Ospelkaus:10a,Park:17,Blackmore:20b}. The different nuclear spin sublevels can be resolved by current techniques if the energy separation  between them is 100 Hz or more \cite{Park:17,Tobias:22}. Thus, the preparation of  the  initial superposition $|\tilde{0}0,+\rangle$  (in {\it Step 1}) and its deexcitation back to the ground rotational state  $|\tilde{0}0,+\rangle$ (in {\it Step 3}) can be realized with high fidelity in current experiments. 

\item
 The ability to time evolve the initial state $|+\rangle$ under the Ising and XXZ Hamiltonians on the timescale of tens of milliseconds without significant loss of coherence.

A detailed analysis of the detrimental effects of decoherence on the entangled states is provided in Sec.~III above. As shown there, these  effects are minor if the coherence times can be made several times longer than the required evolution times. 

The task of maintaining rotational coherence for {much longer than} $2.95$~ms required to realize the cluster state of KRb molecules, is slightly beyond current experimental capabilities. State-of-the-art rotational coherence times of 
$6{-}10$~ms have been  achieved experimentally for KRb {using state-insensitive (magic) trapping techniques \cite{Neyenhuis:12,Tobias:22,Li:23,Guan:21}.}
With recent advances in such techniques, one can reasonably expect to extend these times to 470 ms or longer, as recently shown for CaF molecules in optical tweezers \cite{Burchesky:21}. In addition, increasing the molecular dipole moment  (i.e. using NaCs instead of KRb) can shorten the required dynamical evolution times by several orders of magnitude (because $J_z\simeq d^2$),  significantly mitigating the requirements on the decoherence time.

Additionally,  dynamical decoupling techniques can be used to extend the coherence times. These techniques have been  successfully applied in previous 
experimental work with lattice-confined KRb molecules \cite{Yan:13}.

\end{enumerate}

\vskip9pt
{\bf Experimental requirements for  spin-squeezed state generation  with $^2\Sigma$ molecules }
\vskip3pt

For the spin-squeezed states generation protocol  using  $^2\Sigma$ molecules, we need the following experimental capabilities, in addition to those  listed above.

First, we need to be able to magnetically tune the opposite-parity levels of YO near an avoided level crossing (ALC)
 of the opposite-parity spin-rotational states.  
As shown in Fig. 4(b) of the main text, the ALC in YO($^2\Sigma$) is about 10~G wide, and occurs at $B_c=8,590$ G at 5 kV/cm.
While it is very challenging to experimentally realize such a large  magnetic field, it has been achieved for  BaF$(^2\Sigma)$ molecules ($B_c\simeq 5,000$~G) in a molecular beam experiment \cite{Altuntas:18} using superconducting magnets. 

The required magnetic field can be dramatically reduced by choosing a heavier molecule with a smaller rotational constant, since $B_c\simeq 2B_e/\mu_0$. As an example, consider the SrI($^2\Sigma$) molecule  \cite{Asnaashari:22} recently suggested as a potential candidate for laser cooling \cite{Liu:21}. The rotational constant of SrI is $\simeq$10 times smaller than that of YO \cite{Liu:21,Schroder:88}, reducing the magnetic field  required to reach an ALC to just 815~G, which is  much more amenable to experimental realization.

Second, we need to drive single-photon microwave transitions between the opposite-parity levels near the ALC.  This is possible because the levels have a nonzero transition dipole moment ($d_{\uparrow\downarrow}\ne 0$) as shown in Fig. 4(c) of the main text, and the experimental techniques for coherent microwave manipulation of ultracold molecules are well-established \cite{Ospelkaus:10a, Park:17,Blackmore:20b}. 

Third,  the magnetic field should be switchable on a timescale that is much faster than the optimal squeezing time (see above), i.e., $t_\text{switch}\ll t_0$, where $t_0=0.64$ ms for YO).  This is presently a {significant}  experimental challenge, which could be mitigated by using a smaller value of the crossing field for heavier molecules (see above).

\vskip9pt
{\bf Experimental requirements for  spin-squeezed state generation  with microwave-dressed states of   $^1\Sigma$ molecules  }
\vskip3pt

In this protocol, the squeezed state is created in the mw-dressed basis. Thus,  in addition to conditions (1)-(4) listed above, we need to be able to prepare a long-lived microwave-dressed state with a long coherence time. The optimal squeezing time for mw-dressed KRb molecules at $B=B_c$ is $t_0\simeq 2/J^{\perp}_{i,i+1} =5.3$ ms for $J^{\perp}_{i,i+1}=60$~Hz (see Fig. 5(b) of the main text), and thus we require $T_2\gg 5.3$~ms.

While  mw-dressed states with lifetimes longer than 500 ms have recently been prepared experimentally \cite{Anderegg:21}, to our knowledge the coherence times of superpositions between mw-dressed and bare states are yet to be measured.  These coherence times will be limited by the coherence properties of the dressing mw field, as well by anisotropic (tensor) Stark shifts of the bare rotational state components of the mw dressed state.   Our preliminary calculations suggest that the magic conditions can be realized for mw-dressed states of KRb molecules in a dc  electric field $E>23$~kV/cm \cite{Tscherbul:22b}, which will enable the experimental realization of long coherence times.

Finally, we need to ensure that the  dressing mw field can be adiabatically ramped down on a timescale shorter than the evolution time to convert the mw-dressed qubit into the bare state $|\tilde{1}0,M\rangle$. This operation will leave the molecules in a superposition of bare spin states  $|\tilde{1}0,M\rangle$ and $|\tilde{1},-1,M'\rangle$ with $M\ne M'$ (see the main text), which is not affected by the electric dipolar interaction.  As such, the coherent evolution will be stopped after the optimal spin squeezing time $t_0$, leaving a long-lived spin-squeezed ensemble of KRb molecules.   The adiabatic ramp-down of the dressing mw field has been achieved experimentally on a timescale of 0.1 ms \cite{Anderegg:21}, which is much shorter than the spin squeezing time $t_0=5.3$ ms.

\vskip9pt
{\bf Spin detection efficiency  }
\vskip3pt

Molecules are more challenging to detect than atoms, and a limited detection efficiency would degrade the observed spin squeezing.  
We normally transfer ground state molecules in a coherent fashion to a Feshbach state, and then perform atom detection.  The main limitation in the process arises from the transfer efficiency of both the STIRAP (coherent Raman transfer) and dissociation of Feshbach molecules.  A combined efficiency of 80\% is achievable, indicating that detection of a few dB spin squeezing is possible now.  

\color{black}

\end{document}